\documentclass[revtex,reprint,twocolumn,showpacs,superscriptaddress]{revtex4-1}
\usepackage{amsmath,amssymb,bm,mathrsfs,graphicx, braket, times}
\usepackage{graphicx}
\usepackage[colorlinks=true,citecolor=blue,linkcolor=blue]{hyperref}
\usepackage{epsfig}
\usepackage{bm}
\usepackage{natbib}

\let \oldhat \hat

\renewcommand{\vec}[1]{\bm{#1}}
\renewcommand{\hat}[1]{\oldhat{\bm{#1}}}

\begin{document}
\title{Phonon analogue of topological nodal semimetals}
\author{Hoi Chun Po, Yasaman Bahri, and Ashvin Vishwanath}
\affiliation{Department of Physics, University of California, Berkeley, California 94720, USA}

\begin{abstract}
Topological band structures in electronic systems like topological insulators and semimetals give rise to highly unusual physical properties. Analogous topological effects have also been discussed in bosonic systems, but the novel phenomena typically occur only when the system is excited by finite-frequency probes. A mapping recently proposed by Kane and Lubensky [Nat. Phys. {\bf 10}, 39 (2014)], however, establishes a closer correspondence. It relates the zero-frequency excitations of  mechanical systems to topological zero modes of fermions that appear at the edges of an otherwise gapped system. Here we generalize the mapping to systems with an intrinsically gapless bulk. In particular, we construct mechanical counterparts of topological semimetals. The resulting gapless bulk modes are physically distinct from the usual acoustic Goldstone phonons, and appear even in the absence of continuous translation invariance. Moreover, the zero-frequency phonon modes feature adjustable momenta and are topologically protected as long as the lattice coordination is unchanged. Such protected soft modes with tunable wavevector may be useful in designing mechanical structures with fault-tolerant properties.
\end{abstract}

\maketitle

\section{Introduction}
Correspondences between bosonic and fermionic problems have long been an indispensable tool for both solving and understanding model systems. 
Other than a few known exceptions, however, conventional mappings like the Jordan-Wigner transformation and bosonization are generally exact only in one spatial dimension \cite{Sachdev2011,Wen2004}. 
For free particles, both bosonic and fermionic Hamiltonians are characterized by the single-particle Hamiltonians. The statistics of the particles are reflected in the different algebras invoked in their canonical transformations \cite{vanHemmen1980}.
In addition, for stable bosonic system the single-particle Hamiltonian is required to be positive semi-definite \cite{Gurarie2002,Gurarie2003}. 
Despite such differences, one can nonetheless establish an equivalence between the spectra of a bosonic and a fermionic problem by factorizing the corresponding matrices~\cite{Kane2014,Gurarie2002,Gurarie2003}.
Although this approach is limited to free or weakly interacting particles, unlike most other mappings, for certain problems it is manifestly local in any dimensions. 
It is therefore of interest to explore the extent to which this correspondence can provide a new perspective.

The recent realization of the existence of surface modes dictated by topological properties of the bulk has generated immense interest across the community \cite{TI2013,Wan2011,Burkov2011,Vitelli2014,Chen2014,Paulose2014,Kane2014}. Topological insulators, for instance, are electronic systems that host protected conducting surface states but are insulating in their bulk \cite{TI2013}. Within the free fermion description, the existence of these robust surface states can be predicted, via the bulk-boundary correspondence, by computing the topological invariants associated with the occupied Bloch bands \cite{TI2013}. More recently, certain mechanical (spring-mass) problems were independently found to host robust zero-energy modes localized at boundaries \cite{Sun2012}. The boson-fermion mapping proposed by Kane and Lubensky~\cite{Kane2014} demonstrates that these phenomena are the two sides of the same coin.
In particular, the mapping naturally relates the ground state properties of a phonon problem (i.e.~the spectrum in the zero-frequency limit) to that of a particle-hole symmetric fermionic problem at half-filling, for which the Fermi energy is naturally pinned to zero.

In Ref.\,\cite{Kane2014}, the boson-fermion mapping was applied to mechanical frames with a gapped spectrum (aside from the acoustic modes), and these systems can be viewed as the phonon analogues of weak topological insulators.  The topological nature of the bulk spectrum is reflected as zero-frequency modes or states of self-stressed localized at the boundaries \cite{Kane2014} or defects \cite{Paulose2014}. 
A natural step forward is to identify phonon analogues of topological nodal semimetals (TNS), which feature gapless nodes in the bulk spectrum protected by nontrivial band topology, and can lead to interesting surface Fermi-arcs~\cite{Oskar2014}.

Previous works have demonstrated the construction of TNS analogues in bosonic systems, such as photonic crystals \cite{Lu2013}, acoustic systems~\cite{Meng} and even spring-mass models~\cite{Wang2015}. The associated band touchings, however, occur at non-zero frequencies and the topological features are irrelevant when one is interested in only the low energy excitation of the system. 

In this work, we seek to construct phonon analogues of TNS with protected zero-energy modes in the linearized phonon spectrum.
In particular, we consider pinned, periodic spring-mass models. 
One may expect such models, if stable and not `floppy' (having an extensive number of zero modes), should generically have a fully gapped phonon spectrum at zero phonon frequency. Contrary to this expectation, we show that bulk node with $\omega (\vec k_c) = 0$ can in fact appear without fine-tuning in certain isostatic models, and their existence is rooted in the topological protection of the corresponding fermionic TNS.
These systems correspond to metamaterials or mechanical structures with robust extended soft modes which can be employed as building blocks of more complex structures requiring both rigidity for stability and flexibility for functionality \cite{Chen2014,Paulose2014}. 
Fault tolerance, provided by topological protection, is highly desirable for applications in which different mechanical parts are coupled to perform nontrivial maneuvers. Similarly, our construction can also be applied to engineer acoustic or mechanical metamaterials with programmable response to external excitations \cite{Craster2012,Florijn2014}.

Before we move on to present our results, we pause to comment on the boson-fermion correspondence discussed here.
For non-interacting problems, both fermionic and bosonic Hamiltonians are characterized by a single-particle Hamiltonian $\mathcal H$, and the ground state (zero-temperature) behavior of the system is determined by the `filling' of the eigenmodes of $\mathcal H$. For fermions, Pauli exclusion dictates that all the eigenmodes up to the Fermi energy are filled in the zero-temperature limit. In particular, a 2D nodal semi-metal results if the Fermi surface consists entirely of nodal points, as in graphene (assuming spin-rotation invariance). The excitations about the ground state then correspond to quasi-particles with linear dispersion about some characteristic momentum $\vec k_c$.

Bosons in the zero-temperature limit, however, only occupy the lowest energy single-particle state(s) in the absence of interactions. Therefore, the stability of a free bosonic system requires $\mathcal H$ to be positive semi-definite. As such, one can write $\text{eig}(\mathcal H) = \{ \omega^2\}$ with $\omega \geq 0$, and the ground state is gapped if and only if $\omega^2_{\text{min}}>0$. A non-interacting 2D bosonic analogue of a semimetal will then be a system for which $\omega^2(\vec k_c) = 0$ at a collection of nodal points $\{ \vec k_c\}$ (in the thermodynamic limit), since they would feature similar low-energy excitations.
Strictly speaking, even the acoustic phonons satisfy the criterion outlined, but they are non-topological and can be trivially gapped out by breaking transition invariance (via, say, the introduction of a pinning potential). In contrast, here we are interested in the phonon analogue of \emph{topological} nodal semimetals, where, in contrast to acoustic modes arising from conventional continuous symmetry breaking, the existence of the low-energy excitations are dictated by an underlying topological invariant.

\begin{figure}[bth!]
\centering
\includegraphics[width=0.95 \linewidth]{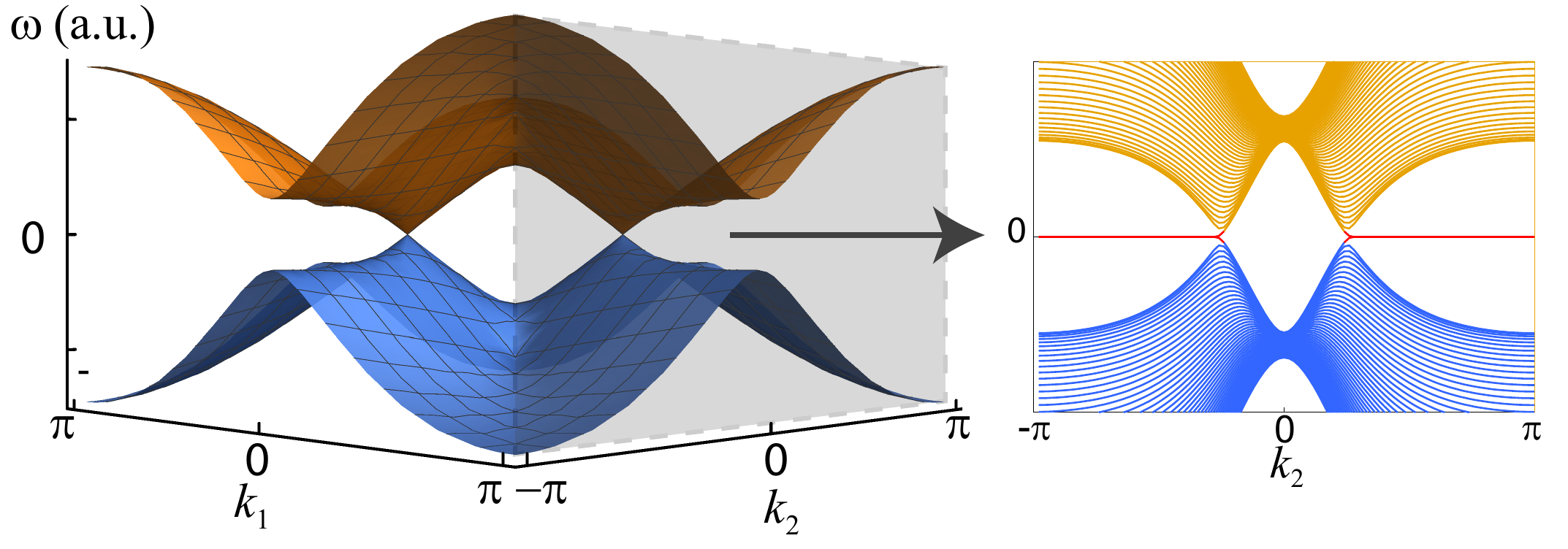} 
\caption{Typical spectrum of a 2D TNS protected by chiral symmetry hosting a pair of topological nodes at $\omega = 0$ (only the two bands close to $\omega = 0$ are shown). The spectrum is computed for the fermionic problem defined in Eq.\,(\ref{eq:FermionicH}) with $R$ defined in Eqs.(\ref{eq:Rk}) and (\ref{eq:SmoothPegvw}). The parameters $(\theta_1,\phi_1,\theta_2,\phi_2) = (1/4,0,1/8,-1/8)\pi$ are used. The inset shows the edge spectrum when open and periodic boundary conditions are enforced for the $\hat e_1$ and $\hat e_2$ directions respectively, featuring a line of zero modes connecting the projections of the bulk topological nodes onto the surface Brillouin zone.
\label{fig:FermionicSpec}
}
\end{figure}

\section{Phonon analogue of topological nodal semimetals}
For a self-contained discussion, we first briefly describe phonon problems defined for mechanical structures formed by mass points connected by elastic elements modeled as central-force springs, and we review the boson-to-fermion mapping developed in Refs.\,\cite{Kane2014,Gurarie2002,Gurarie2003}. 
The dynamics of such a system is determined by the kinetic energy of the mass points and the elastic potential energy stored in the springs, and therefore for $N$ masses in $d$ spatial dimension, a generic phonon problem can be defined by the Hamiltonian
\begin{equation}\begin{split}\label{eq:PhH}
H_B =& \sum_{\vec r}  \left ( \frac{\vec p_{\vec r}^2 }{2m_{\vec r}} + \frac{1}{2} \sum_{j=1}^{z/2} \kappa_{j, \vec r} s_{j, \vec r}^2  
\right),
\end{split}\end{equation}
where $\vec r$ denotes the equilibrium positions of masses $m_{\vec r}$, $\vec p$ denotes momentum of the masses, $z$ is the coordination number, and  $s_{j, \vec r}$, $ \kappa_{j, \vec r}$ are respectively the extension and spring constant of the $j$-th spring of the unit cell at $\vec r$. To simplify discussion, we set both $m_{\vec r}$  and $\kappa_{j, \vec r}$ to unity for all $(j, \vec r)$ unless otherwise specified. Since we are interested in the existence of bulk zero modes, we also impose periodic boundary conditions.

The spring extensions $\{ s_{j, \vec r} \}$ are functions of the displacements $\{ \vec x_{\vec{r}} \}$ of the masses they connect. Within the harmonic approximation, the extensions are expanded to linear order in the mass displacements: $\vec S= R \vec X + \mathcal O(\vec X^2)$, where $\vec X$ is a $d N$ dimensional column vector aggregating the displacement vectors of the $N$ masses, and $\vec S$, similarly defined for the extensions $\{ s_{j, \vec r} \}$, is $zN/2$ dimensional. The $zN/2\times dN$ dimensional matrix $R$ is physically a linear map relating the spring extensions to the mass displacements. $R^T$ is known as the equilibrium matrix and it relates the acceleration of the masses to the spring extensions: $\ddot {\vec X} = - R^T \vec S$. As in \cite{Kane2014}, henceforth we restrict attention to isostatic lattices, which have  $z = 2d$ and so $R$ is a square matrix. With this notation, the phonon Hamiltonian, within the harmonic approximation, can be recast as $ H_B= \frac{1}{2} \left[  \vec P^2+  \left( R  \vec X \right)^2\right]$, where $\vec P$ is similarly defined as $\vec X$ and $R^T R=D$  is the real-space dynamical matrix. Equivalently, one can view $R$ as a factorization of $D$.

The phonon modes are solutions to the eigenvalue problem $D \xi_i = R^T R \xi_i = \omega_i^2 \xi$, where $\omega_i$ is the eigenfrequency of the $i^{th}$ mode with $i = 1,\dots, dN$. The bosonic phonon problem, characterized by $R$, can be mapped to a fermionic problem by considering a chiral matrix $\mathcal H_F$ and the associated Hamiltonian $H_F$ \cite{Kane2014,Gurarie2002,Gurarie2003}
\begin{equation}\begin{split}\label{eq:FermionicH}
\mathcal H_F = \left(
\begin{array}{cc}
0 & - i R^T\\
i R & 0
\end{array}
\right);~ H_F = \left(
\begin{array}{cc}
\bar {\vec \chi }& \vec \chi
\end{array}
\right)
\mathcal H_F
\left(
\begin{array}{c}
\bar {\vec \chi }\\
 \vec \chi
\end{array}
\right).
\end{split}\end{equation}
$\mathcal H_F$ satisfies $\{ \tau^z , \mathcal H_F\} = 0$ with $\tau^z = \text{diag}(1_{dN\times dN},-1_{dN\times dN})$. One can easily verify the $2dN$ eigenvalues of $\mathcal H_F$ are given by $\{\pm \omega_i \}$, which implies the phonon spectrum is encoded in the energy spectrum of the fermionic Hamiltonian $ H_F$. We note that only a subset of the zero modes of $\mathcal H_F$ gives rise to zero modes of $D$; namely, those fermionic modes with $\tau^z = 1$. In contrast, fermionic zero-energy modes with $\tau^z = -1$ (i.e. null vectors of $R^T$) give rise to the so-called states of self-stress of the mechanical problem \cite{Kane2014}. $\vec \chi, \bar{\vec  \chi}$ are Majorana fermions satisfying $\{ \chi_l,\chi_m\} = \{ \bar  \chi_l ,\bar \chi_m \}= \delta_{lm}$ with all other anti-commutators vanishing. In particular, $\chi$ ($\bar \chi$) is even (odd) under time reversal (TR). Since $ H_F$ corresponds to a TR symmetric Hamiltonian of spinless fermions, it belongs to the Altland-Zirnbauer symmetry class BDI \cite{AltlandZirn}.

As the corresponding fermionic Hamiltonian is chiral, it can host topological nodes without fine-tuning even in 2D. To find a phonon analogue inheriting the topological features of the TNS, therefore, we consider the phonon spectrum arising from a spring-mass model defined on the square lattice ($d=2$ and $z=4$). 
For a regular phonon problem, the form of $R$ is constrained by global continuous translation invariance and spatial symmetries of the underlying lattice. For the square lattice, one simply finds $\omega_{j}(\vec k) \propto | \sin k_j|$, where $j=1,2$ labels the two orthogonal directions. Although $\omega_j (k_j = 0) = 0$, these nodal lines are not topologically protected as they can be gapped out by breaking translation symmetry.

\begin{figure}[t]
\centering
\includegraphics[width=0.7 \linewidth]{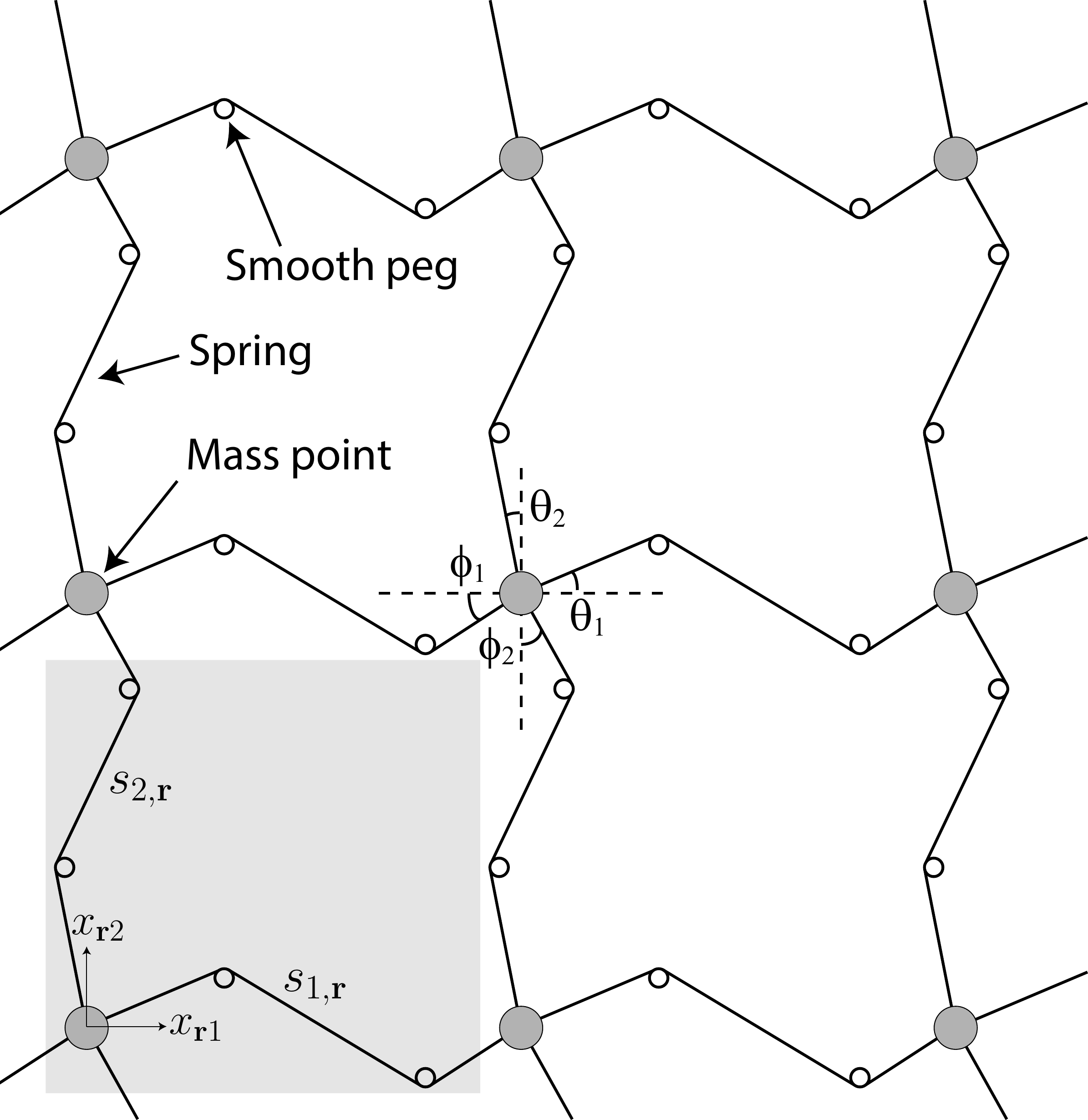} 
\caption{A possible realization of the considered phonon problem. The dependence of the spring extensions on the displacements of the masses can be modified by bending the springs with fixed, smooth pegs which serve as a pinning potential and gap out the acoustic modes. The spring extensions are then characterized by Eq.\,(\ref{eq:SmoothPegvw}) to linear order in mass displacements. The shaded region indicates the unit cell convention adopted.}
\label{fig:SmoothPeg}
\end{figure}

Here we relax from these symmetry constraints and assume they are explicitly broken. A possible realization is depicted in Fig.\,\ref{fig:SmoothPeg}, in which the springs are tweaked with fixed, smooth pegs that serve as an external pinning potential (a more realistic model is presented in Appendix A). The form of $R$, however, is still constrained by the geometrical relations between the spring extensions and mass displacements. For a spring connecting the masses at equilibrium positions $\vec r = \vec a$ and $\vec b$, the spring extension should satisfy $|s|\leq | \vec x_{\vec a}| + | \vec x_{\vec b}|$. In particular, we assume there is a special spatial direction $\vec v= \cos \theta_a \hat e_1 + \sin \theta_a \hat e_2$ such that the inequality is saturated for $\vec x_{\vec a} = |\vec x_{\vec a}| \vec v$ and $\vec x_{\vec b} = \vec 0$. Equivalently, this implies $ \partial_{x_{\vec a 1}} s = \cos{\theta_a } $ and $\partial_{x_{\vec a 2}} s = \sin{\theta_a}$  in the original basis. Assuming a similar dependence of $s$ on $\vec x_{\vec b}$, characterized by $\vec w= \cos \theta_b \hat e_1 + \sin \theta_b \hat e_2$, one finds $s = \vec v \cdot \vec x_{\vec a} + \vec w \cdot \vec x_{\vec b} + \mathcal O (\vec x ^2)$.

For a clean system, all springs that are equivalent under lattice translations are characterized by the same parameters and in momentum space one finds
\begin{equation}\begin{split}\label{eq:Rk}
R(\vec k ) = 
\left(
\begin{array}{cc}
v_{11} + w_{11} e^{-i k_1} & v_{12} + w_{12}e^{-i k_1}\\
v_{21} + w_{21}e^{-i k_2} & v_{22} + w_{22} e^{ -i k _2}
\end{array}
\right),
\end{split}\end{equation}
where $\vec k = (k_1, k _2)$ lies in the first Brillouin zone (BZ), and $v_{lm}$ ($w_{lm}$) denotes the $m$-th component of the vector $\vec v_l$ ($\vec w_l$), which relates the spring extension to the displacements $\vec x_{\vec r}$ and $\vec x_{\vec{r}+\hat e_l}$. Enforcing the geometric constraints, the vectors can be parameterized by
\begin{equation}\begin{split}\label{eq:SmoothPegvw}
\vec v_{1} = 
\left(
\begin{array}{c}
-\cos \theta_{1}\\
-\sin\theta_{1}
\end{array}
\right)
;&~ \vec v_{2}= 
\left(
\begin{array}{c}
\sin \theta_2\\
-\cos \theta_2
\end{array}
\right)
; \\
\vec w_{1} = 
\left(
\begin{array}{c}
\cos \phi_{1}\\
\sin \phi_{1}
\end{array}
\right)
;&~
 \vec w_2 =  
 \left(
\begin{array}{c}
-\sin \phi_2\\
\cos \phi_2
\end{array}
\right).
\end{split}\end{equation}
The sign convention of the angles are chosen to match the parameterization in Fig.~\ref{fig:SmoothPeg}. 
The corresponding fermionic Hamiltonian, as defined in Eq.\,(\ref{eq:FermionicH}), is explicitly given by
\begin{equation}\begin{split}\label{eq:HF2}
H_F = 2 i \sum_{\vec r} \sum_{l,m=1}^2&\left(   v_{lm} \chi_{\vec r}^{l} \bar\chi_{\vec r}^{m} + w_{lm} \chi_{\vec r}^{l} \bar \chi_{\vec r + \hat e_l}^m \right).
\end{split}\end{equation}
Although the BDI class is topologically trivial in 2D, the system can inherit the nontrivial properties of the 1D system when identical 1D chains are stacked in one direction, similar to weak topological insulators. To this end, we introduce Fourier-transformed variables
\begin{equation}\begin{split}\label{eq:}
\chi_{\vec r}^l = \frac{1}{\sqrt{N_2}} \sum_{k_2} \chi_{x_1,k_2}^l e^{- i k_2 x_2}; ~\chi^l_{x_1,k_2} = \frac{1}{\sqrt{N_2}} \sum_{x_2} \chi_{\vec r}^l e^{  i k_2 x_2},
\end{split}\end{equation}
where $N_j$ is the number of sites along the $\hat e_j $ direction, and $\bar \chi_{x_1,k_2}^l$ is similarly defined. Since $(\chi_{x_1,k_2}^l )^\dagger= \chi_{x_1,-k_2}^l $ is complex, one can define Majorana operators for $k_2 \in (0,\pi)$
\begin{equation}\begin{split}\label{eq:}
\lambda_{x_1,|k_2|} = \frac{\chi_{x_1, k_2 }^l + \chi_{x_1,- k_2 }^l}{2}; ~~
\eta_{x_1,|k_2|} = \frac{\chi_{x_1, k_2 }^l - \chi_{x_1,- k_2 }^l}{2i},
\end{split}\end{equation}
which are both even under TR. $\bar \lambda_{x_1,|k_2|}$ and $\bar \eta_{x_1,|k_2|} $, similarly defined for $\bar \chi_{x_1, k_2}$, are odd under TR. Since Eq.(\ref{eq:HF2}) only couples $\chi$ to $\bar \chi$, upon Fourier transform $H_F(k_2)$ will only couple $( \lambda ,\eta)$ to $(\bar \lambda , \bar \eta)$  and each such 1D system is in the BDI class. While the BDI class is classified by $\mathbb Z$ in 1D, the indexes of the 1D chains here are always even: the original lattice translational symmetry along $\hat x_2$, reflected as a local $\mathcal O(2)$ rotation between $(\lambda_{x_1,|k_2|}, \eta_{x_1,|k_2|})$, guarantees that the Majorana zero modes at each edge occur in pairs. 
This can also be understood from the doubling of Majorana modes for each $k_2\in(0,\pi)$, as $\chi^l_{x_1,k_2}$ and $\chi^l_{x_1,-k_2}$ are now Hermitian conjugates of each other. 
Such property of the system is more manifest by relabeling the operators as
\begin{equation}\begin{split}\label{eq:}
c_{l,x, k_2}^A \equiv \chi_{x_1, k_2 }^l;&~~~c_{l,x, k_2}^B \equiv \bar \chi_{x_1, k_2 }^l; ~ k_2 \in(0,\pi)
\end{split}\end{equation}
such that $c_l$'s are complex fermions. Eq.(\ref{eq:HF2}) is then decoupled into a series of 1D fermionic Hamiltonians, each labeled by $k_2 \in (0,\pi)$,
\begin{equation}\begin{split}\label{eq:HFk2}
H_F (k_2) = 2   \sum_{ x}  &\left[   
i v_{11} c_{1,x}^{A \dagger} c_{1,x}^B
+ i v_{12} c_{1 ,x}^{A\dagger }c_{2 ,x}^B\right.\\
&\left.
~+i  w_{11} c_{1 ,x}^{A\dagger} c_{1 ,x+1}^B +i w_{12} c_{1,x}^{A\dagger} c_{2,x+1}^B
\right.\\
&\left.
~+i (v_{21} +w_{21} e^{-i k_2}) c_{2,x}^{A\dagger} c_{1,x}^{B} \right.\\
&\left.
~+ i ( v_{22} +  w_{22} e^{-i k_2} ) c_{2 ,x}^{A\dagger }c_{2 ,x}^{B} + h.c.
\right],
\end{split}\end{equation}
where we have suppressed the subscript $k_2$ on the operators. The original chiral symmetry is manifested in this notation as a sublattice (A-B) symmetry. 

To characterize these 1D systems, we further Fourier transform on $x$, which gives
\begin{equation}\begin{split}\label{eq:}
H_F (k_2) = 2 \sum_{k_1}
\left(
\begin{array}{cc}
\vec c^{A \dagger}& \vec c^{B \dagger }
\end{array}
\right)
\left(
\begin{array}{cc}
0 & i R_{\vec k}\\
- i R_{\vec k}^\dagger & 0
\end{array}
\right)
\left(
\begin{array}{c}
\vec c^A\\
\vec c^B
\end{array}
\right),
\end{split}\end{equation}
where the subscript $\vec k$ of the operators is suppressed, $\vec c^{A} = (c^A_{1}, c^A_{2})^T$, and similarly for $\vec c^B$. The bulk topological invariant of $\mathcal H_F(k_2)$ is given by the winding number of $\det(i R_{\vec k})$ as $k_1$ is varied from $-\pi$ to $\pi$ \cite{TI2013}. More explicitly, we have
\begin{equation}\begin{split}\label{eq:}
\det (i R_{\vec k}) = &- \left( [\vec v_1 \wedge \vec v_2] +  [\vec v_1 \wedge \vec w_2]  e^{-i k_2}  \right) \\
&- \left(  [\vec w_1 \wedge \vec v_2]  +  [\vec w_1 \wedge \vec w_2]  e^{-i k_2}  \right)  e^{- i k_1}\\
\equiv& r_1 ( k_2 ) + r_2 ( k_2 ) e^{-i k_1},
\end{split}\end{equation}
where $[\vec v \wedge \vec w ]$ denotes the component of the wedge product $\vec v \wedge \vec w$ in the $\hat e_1 \wedge \hat e_2$ direction, and $r_1(k_2)$, $r_2(k_2)$ are introduced to simplify the expressions. The winding number $\mathcal W_{k_2}$ is determined by the relative magnitude of $ \left | r_1(k_2)\right|$ and $ \left | r_2(k_2)\right|$. The case of particular interest is when the winding numbers $\mathcal W_{ k_2  \rightarrow 0^+} \neq \mathcal W_{ k_2  \rightarrow \pi^-}$, which implies there must be a topological phase transition as $ k_2 $ is changed from $0$ to $\pi$. Such a phase transition occurs when 
\begin{equation}\begin{split}\label{eq:}
\frac{| r_1 (0) r_1(\pi)| + |r_2(0) r_2(\pi)|}{| r_1 (0) r_2(\pi)| + |r_2(0) r_1(\pi)|}<1,
\end{split}\end{equation}
and when this is satisfied the gap at $E=0$ must close at some critical quasi-momentum $\vec k_c$, giving rise to a topological node. Such nodes in the fermionic picture are reflected in the original bosonic problem as a pair of isolated points $\pm \vec k_c$ at which the phonon frequency vanishes, corresponding to a bulk zero mode in the linearized spectrum.

For a system parameterized as in Eq.\,(\ref{eq:SmoothPegvw}), we have
\begin{equation}\begin{split}\label{eq:}
&\frac{| r_1 (0) r_1(\pi)| + |r_2(0) r_2(\pi)|}{| r_1 (0) r_2(\pi)| + |r_2(0) r_1(\pi)|}\\
=&\max\{| \cos{(\theta_1 - \phi_1)}|, |\cos{(\theta_1 + \phi_1 - \theta_2 - \phi_2)}| \},
\end{split}\end{equation}
which implies that, for general parameters, its linearized phonon spectrum always contains a topologically protected bulk zero mode. Note that if $\theta_1 + \phi_1 - \theta_2 - \phi_2 $ is an integer multiple of $\pi$ then $\omega$ accidentally vanishes on a pair of arcs in the BZ, rendering $\mathcal H_F(|k_2|)$ gapless for all $|k_2|$. For generic parameters, the critical quasi-momentum is 
\begin{equation}\begin{split}\label{eq:}
\vec k_c = 
(\theta_1 - \phi_1) \hat x + (\theta_2 - \phi_2) \hat y,
\end{split}\end{equation}
and therefore the quasi-momentum associated with the protected bulk zero mode can be adjusted simply by tuning the angles $\theta_j$ and $\phi_j$.

\section{Dispersion and robustness of zero modes}
\subsection{Conical phonon dispersion for the smooth peg model}

\begin{figure}[b]
\begin{center}
\includegraphics[width = 0.8 \linewidth]{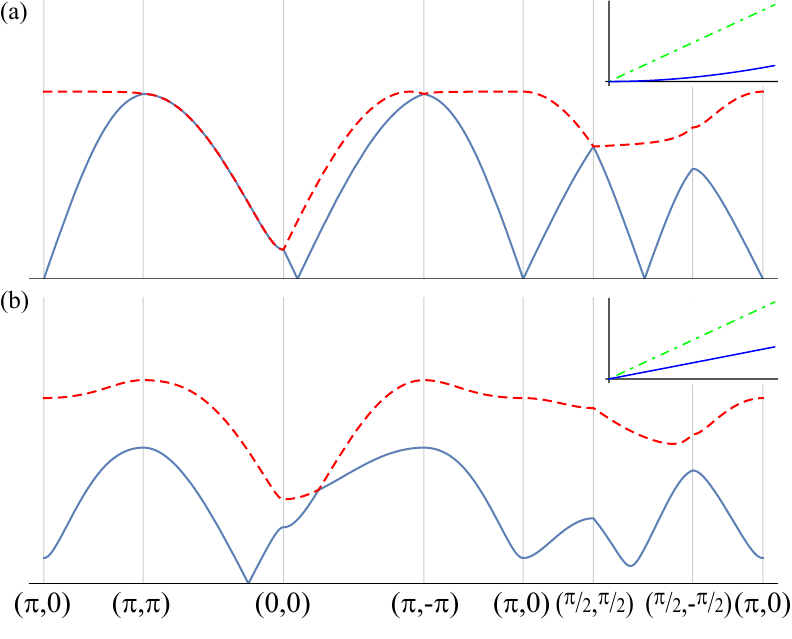}
\caption{
Linearized phonon spectrum for the smooth peg model along different paths in the 2D BZ. The insets show the dispersion of the two independent phonon modes around the critical quasi-momentum $\vec k_c$. As in the text here we take $m=\kappa_1=\kappa_2 = 1$. (a) The parameters $(\theta_1,\phi_1,\theta_2,\phi_2) = (0.1, 0.2, 0.2, 0.1)\pi$ are used. As such $\theta_1+\phi_1 - \theta_2 - \phi_2=0$, which gives rise to an accidental vanishing of $\omega$ along a pair of arcs in BZ. This is seen in the vanishing of $\omega$ at multiple values of $\vec k$ as well as the existence of a quadratic phonon mode around $\vec k_c$. (b) The parameters $(\theta_1,\phi_1,\theta_2,\phi_2) = (1/4, 0, 1/8, -1/8)\pi$ are used (same as Fig.~\ref{fig:FermionicSpec} of the main text), which corresponds to the generic case in which the linearized phonon spectrum contains a pair of isolated bulk nodal points at $\pm \vec k_c$ with conical dispersion around them.
}
\label{fig:SmoothPegSpect}
\end{center}
\end{figure}

For a stable system, the phonon frequencies satisfy $\omega^2_{\pm}(\vec k) \geq 0$, where the $\pm$ sign corresponds to the two branches from diagonalizing the $2 \times 2$ dynamical matrix $D_{\vec k}$. As such $\omega_{-}^2 (\vec k)$ attains a global minimum at $\vec k_c$ and therefore $\vec \nabla_{\vec k} \omega_{0} |_{\vec k_c} = \vec 0$. Expanding $\omega_{-}^2(\vec k)$ around the nodal point $\vec k_c$, we have 
\begin{equation}\begin{split}\label{eq:ConicalD}
\omega_{-}^2 (\vec k_c + \delta \vec k) \approx \frac{1}{2} \sum_{ij} \left. \frac{\partial^2 (\omega_{-}^2) }{\partial k_i \partial k_j} \right|_{\vec k_c}  \delta k_i \delta k_j.
\end{split}\end{equation}
The phonon speeds around the conical dispersion is given by:
\begin{equation}\begin{split}\label{eq:}
\frac{1}{2}
 \left. \left(\frac{\partial^2 (\omega_{-}^2) }{\partial k_i \partial k_j} 
\right)\right|_{\vec k_c}  &=
\frac{\kappa_1 \kappa_2}{2 (\kappa_1 \mathcal S_1^2 + \kappa_2 \mathcal S_2^2 )} 
\left(
\begin{array}{cc}
\mathcal S_2^2 & \mathcal S_1 \mathcal S_2 \mathcal C_\delta \\
\mathcal S_1 \mathcal S_2 \mathcal C_\delta& \mathcal S_1^2
\end{array}
\right)
\end{split}\end{equation}
where we let $\mathcal S_j = \sin k_{cj}$ with $k_{c j } = \theta_j - \phi_j$ for $j=1,2$, and $\mathcal S_\delta = \sin \delta$, $\mathcal C_\delta = \cos\delta$  with $\delta = \theta_1 + \phi_1 - \theta_2 - \phi_2$. The characteristic speeds around the nodal point is therefore given by 
\begin{equation}\begin{split}\label{eq:}
c^2_{\pm} =&
\frac{\kappa_1 \kappa_2 (\mathcal S_1^2 + \mathcal S_2^2)}{4(\kappa_1 \mathcal S_1^2 + \kappa_2 \mathcal S_2^2)}\left( 1 \pm \sqrt{ 1-   \left( \frac{2 \mathcal S_1  \mathcal S_2 \mathcal S_\delta }{ \mathcal S_1^2 + \mathcal S_2^2  } \right)^2}\right).
\end{split}\end{equation}
It is clear that when $\delta = 0$, we have $c_- = 0$, corresponding to the softening of the phonon mode due to the line node; when $\delta = \pi/2$ and $\mathcal S_1^2 = \mathcal S_2^2$, we have $c_+ = c_-$ and this gives an isotropic conical dispersion.

\subsection{Robustness}
\begin{figure}[htb]
\begin{center}
\includegraphics[width = 0.95 \linewidth]{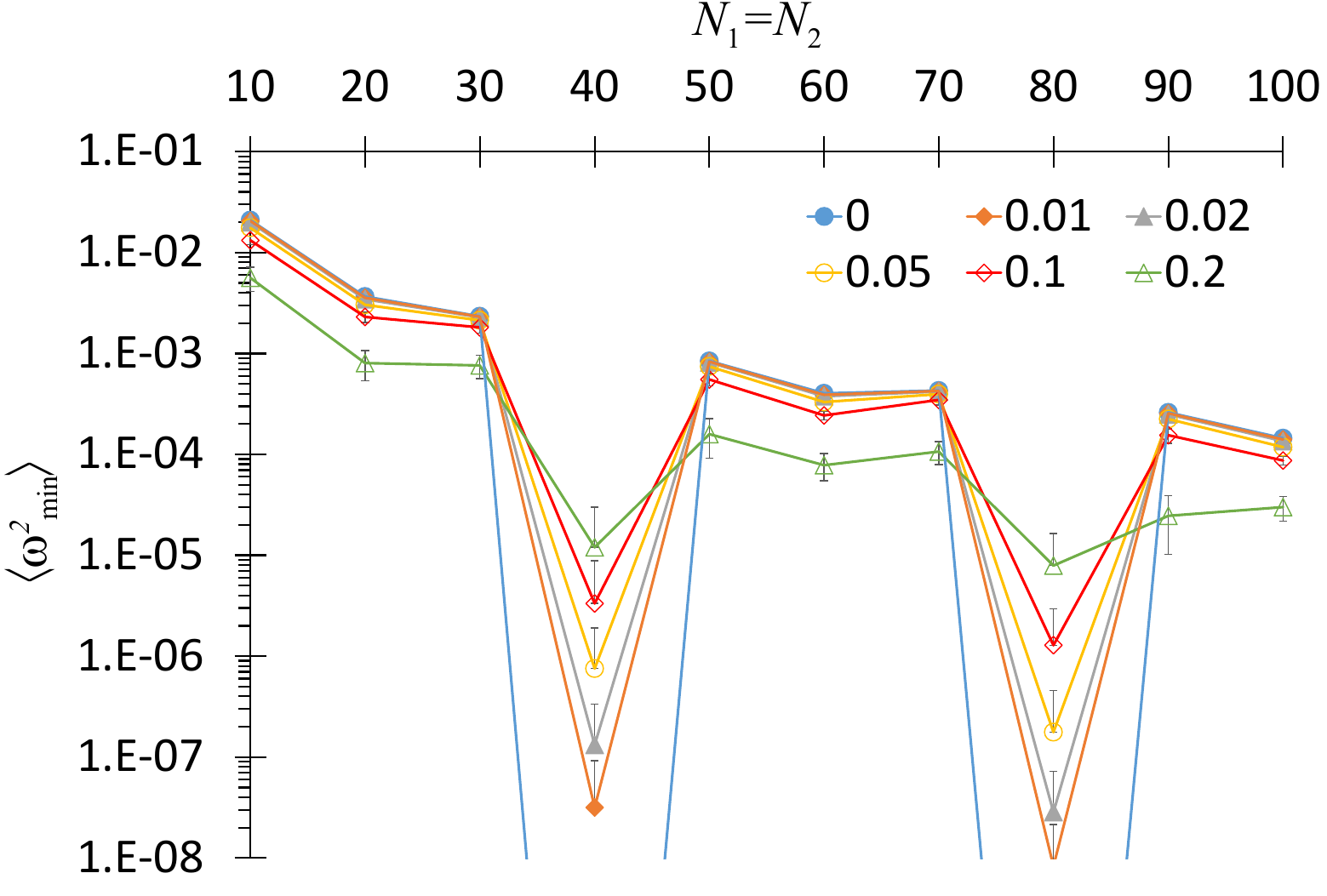}
\caption{
Finite size scaling of the lowest eigenvalue of the disordered dynamical matrix. The linearized dynamical matrix for systems with size $N=N_1 N_2$ and site-dependent couplings is numerically diagonalized to find the smallest eigenvalue $\omega^2_{\text{min}}$. The clean system is defined by the same set of parameters used in Fig.~\ref{fig:FermionicSpec}. Disorder is incorporated by adding to each parameter at each site a small, independent deviation drawn from a normal distribution with standard deviation $\sigma$ ($\kappa$ and $m$ are kept non-negative by taking absolute values). For each $\sigma$, the mean value $\langle \omega^2_{\text{min}}\rangle$ is obtained from $200$ disorder realizations. Error bars represent standard deviation in the disorder averages, and solid lines are guides for the eye. 
\label{fig:DisAvg}
}
\end{center}
\end{figure}

Since the two topological nodes located at $\pm \vec k_c \neq \vec 0$ are isolated in momentum space, they are robust, in the fermionic picture, against small perturbations respecting chiral symmetry. This implies the phonon analogue is robust against weak arbitrary perturbations to the rigidity matrix $R$, accommodating all the natural perturbations in a spring-mass model. Such robustness is demonstrated in the finite-size scaling shown in Fig.\,\ref{fig:DisAvg}, in which we evaluate the disorder average of the lowest eigenfrequency found by numerically diagonalizing the linearized dynamical matrix. The sharp dips at $N=40^2$ and $80^2$ originate from the commensuration between the finite momentum mesh and $\vec k_c = \pm (1/4,1/4) \pi$.
In the presence of disorder ($\sigma \neq 0$) the quasi-momentum ceases to be a good quantum number, but the bulk zero mode remains, as demonstrated by the decrease of $\langle \omega^2_{min}\rangle$ as system size increases.

\begin{figure}[tbh]
\begin{center}
\includegraphics[width = 0.95 \linewidth]{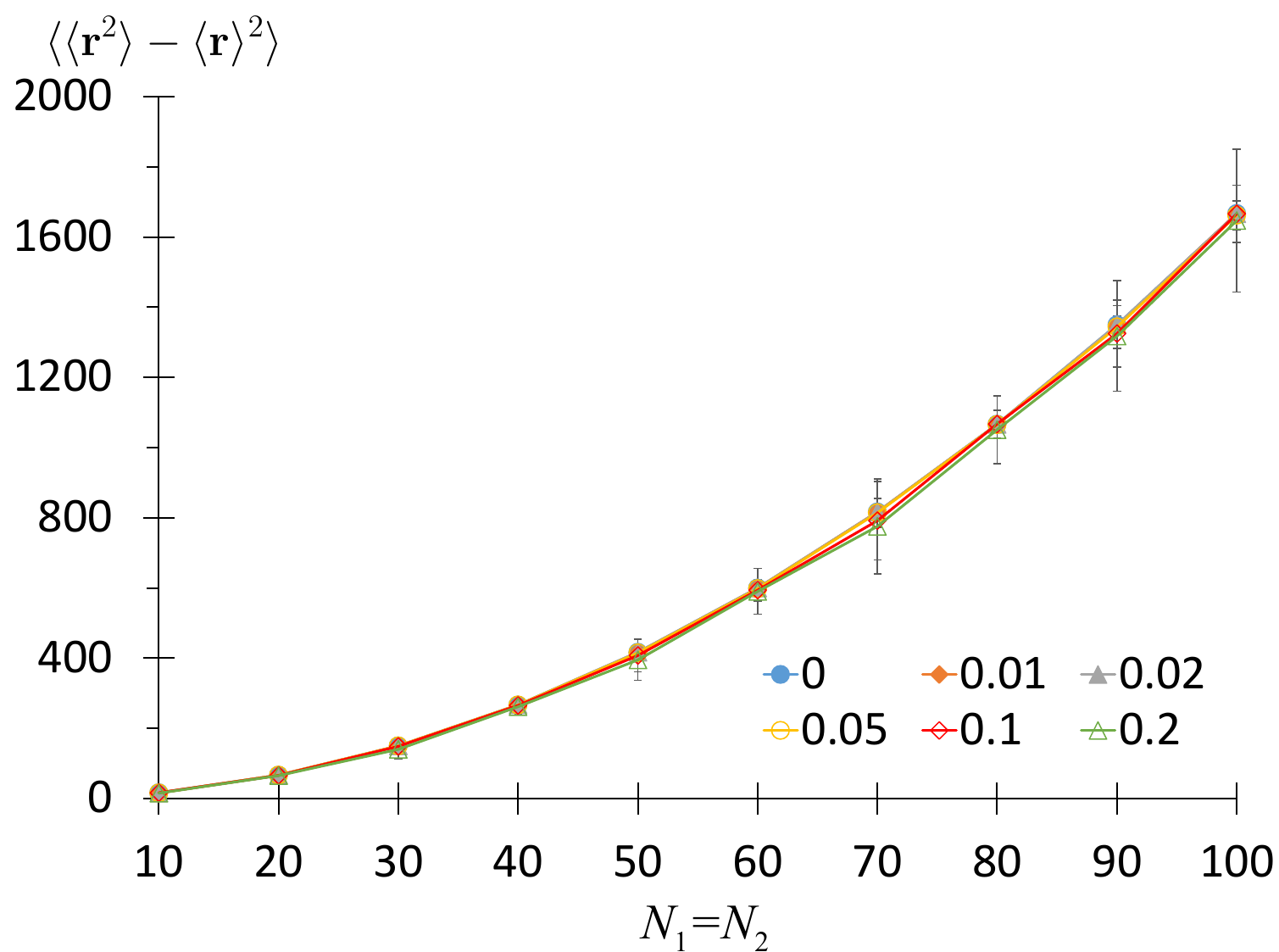}
\caption{
Disorder average of the spatial profile of the lowest-frequency mode. Error bars indicate standard deviation from the disorder averaging.
\label{fig:StateVarAvg}
}
\end{center}
\end{figure}

In the presence of disorder, however, one can reasonably question whether the zero mode observed is really an extended bulk modes, or corresponds to local "rattlers". In Fig.~\ref{fig:StateVarAvg} we plot the disorder average of the variance of the center-of-mass for the state found. The various curves with different disorder strengths collapse to the disorder-free one, indicating the mode remains extended in nature for the disorder strengths considered. 

Since the extended soft modes demonstrated here are protected by the topological properties of the linearized problem, one expects these modes to be only infinitesimal instead of finite. In addition, when the spring-mass description is only a model for a more general system, natural perturbations like the inclusion of small further neighbor couplings correspond to perturbations to the dynamical matrix instead of the rigidity matrix. Such perturbations can render the problem at hand non-isostatic and completely alter the structure of the analysis. The topological protection of the modes is generally lost when confronted with such perturbations. Therefore, we distinguish between the usefulness of this mapping when the bosonic problem is interpreted in a mechanical context (as done here), where local isostaticity is a binary question, as opposed to a quantum mechanical problem obtained after quantization, for which local isostaticity may only be an approximation~\cite{Lawler}. 

\section{Conclusions}
In summary, we show that the mapping of bosonic phonon problems to chiral fermionic problems can be used to construct phonon analogues of topological nodal semimetals. In particular, we construct a 2D system that hosts a tunable bulk extended mode in its linearized phonon spectrum even when global translation symmetry is broken by an external pinning potential. Contrary to usual collective phonon modes arising from continuous symmetry breaking, the existence of such modes is not dictated by symmetry and has its roots in the topological properties of the corresponding fermionic problem. 

Such gapless topological bosonic modes are robust against disorder within the linearized description. These modes should be contrasted with the conventional acoustic modes, which have a completely different origin and are non-topological in nature. It is also worthwhile to point out that disordered, globally isostatic mechanical frames can nonetheless contain over- and under-coordinated regions, and the under-coordinated regions will in general feature localized, `rattling' zero modes. In contrast, the topological modes discussed in this work remain extended even when disorder is introduced.
These topological modes, however, are expected to be gapped when one performs a full phonon spectrum analysis incorporating anharmonicity. As long as the harmonic approximation is justified, the real phonon spectrum still contains such bulk soft modes at finite wavevector and they can be utilized to engineer metamaterials and mechanical structures with fault-tolerant properties.

In finalizing this manuscript, we became aware of a complementary investigation~\cite{Zeb} by Rocklin \emph{et al.} addressing the same problem from a different perspective.

\emph{Acknowledgment} - We thank Haruki Watanabe and Anton Akhmerov for discussions. AV was supported by ARO MURI Grant W 911-12-0461. HCP appreciates financial support from the Hellman Fellows Fund; YB acknowledges funding from NSF GRFP under Grant No. DGE 1106400.

\begin{appendix}
\section{An alternative model}

\begin{figure}[htb]
\begin{center}
{\includegraphics[width=0.95 \linewidth]{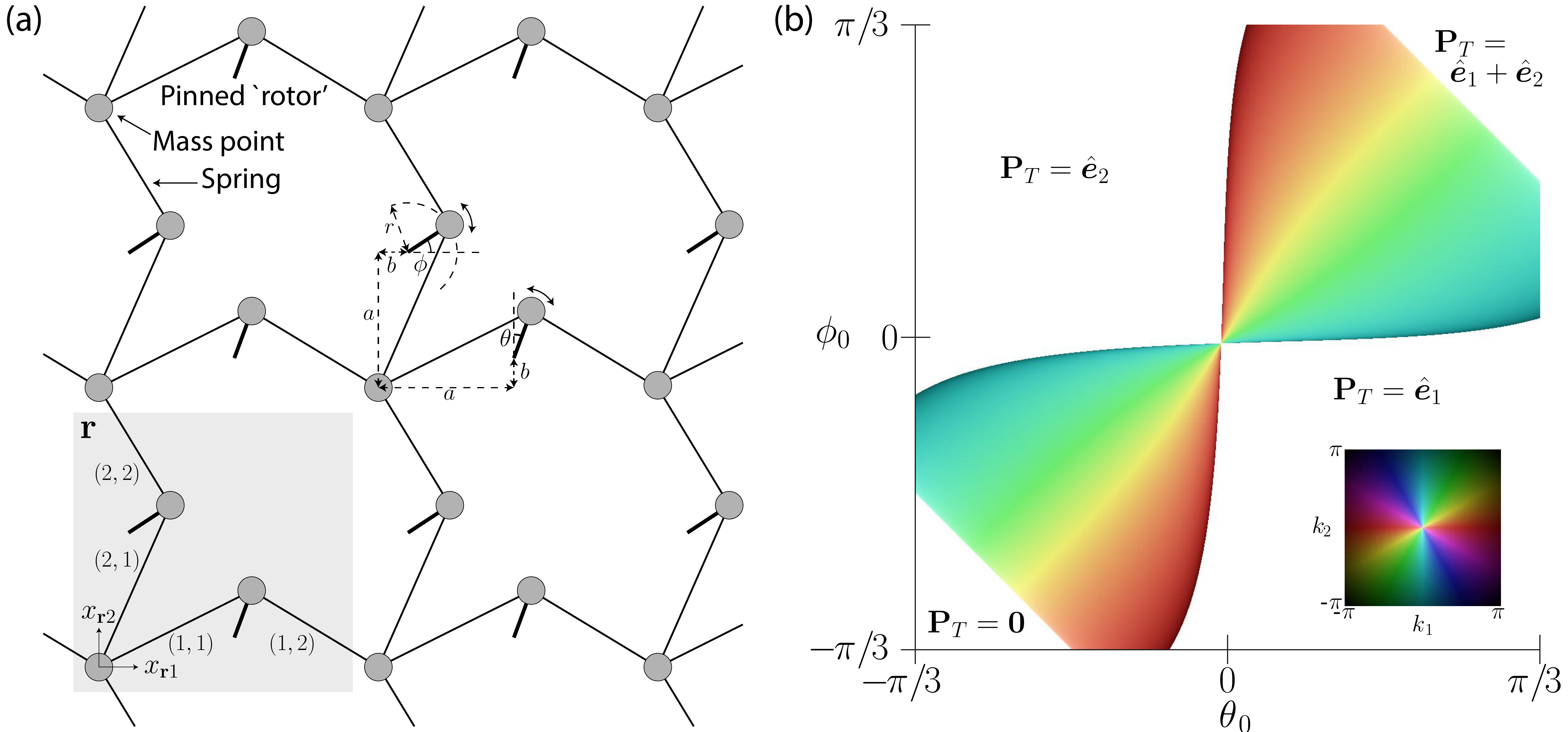}} 
\caption{An alternative model in which the smooth pegs are replaced by pinned rotors, which preserves isostaticity. (a) Schematic of the model indicating the different parameters. The shaded region indicates the choice of the unit cell, and each spring in the cell is labeled by a pair of numbers $(l,m)$. (b) Example phase diagram with $a=0.4$, $b=-0.15$, $r=0.1$ and $\theta_0,~\phi_0 \in [-\pi/3,\pi/3]$. The colored region corresponds to the TNS phase, with the color (inset) encoding the value of $\pm \vec k_c$. The white regions correspond to a gapped phonon spectrum, which can still have different topological polarization $\vec P_{T}$.
\label{fig:aModel}
}
\end{center}
\end{figure}

Here we provide an alternative model that is both more realistic and also exhibits a richer phase diagram. Instead of connecting adjacent mass points by springs sliding over smooth pegs, we considering connecting them via pinned, rigid rotors similar to those discussed in Ref.~\cite{Kane2014}  (Fig.~\ref{fig:aModel}a). 
By `rotor' we refer to a mass point restricted to rotate on a circle by a rigid rod of negligible mass.
This introduces two extra degrees of freedom in each unit cell, but at the same time each of the original spring is replaced by two springs. Altogether the balance between number of degrees of freedom and constraints is maintained, i.e.~isostaticity is preserved.

Each rotor introduced can be characterized by four parameters: mass $m$, radius $r$ of the circle it sweeps out, and the coordinate $(a,b)$ of the pinning point relative to the origin of the unit cell. While these parameters can be different for the two rotors in the unit cell, for simplicity we characterize them by the same set of parameter $(m,r,a,b)$ (see Fig.~\ref{fig:aModel}a). In particular, we set the mass $m$ to be the same as that of the mass points. The spring constants for all springs are also set to the same value $\kappa$. We set the lattice constant to $1$. As such, there are still two system parameters: $\theta_0$ and $\phi_0$, which characterize the tilt of the rotors with
 respect to the coordinate axes at equilibrium. The equilibrium lengths of the springs (assuming no pre-stress), $\ell _{lm}$, can then be written in terms of these parameters:
\begin{equation}\begin{split}\label{eq:}
\ell_{11} =& \sqrt{(b+r \cos \theta_0)^2 + (a+r \sin \theta_0)^2};\\
\ell_{12} =& \sqrt{(b+r \cos \theta_0)^2 + (1-a-r \sin \theta_0)^2};
\end{split}\end{equation}
and $\ell_{21}$, $\ell_{22}$ take the same form as $\ell_{11}$, $\ell_{12}$ but with $\theta_0 \rightarrow \phi_0$.

The energy of the system is 
\begin{equation}\begin{split}\label{eq:}
E =& \frac{m}{2} \sum_{\vec r} \left( \dot x_{\vec r 1}^2+ \dot x_{\vec r 2}^2+  r^2 (\dot \theta_{\vec r}^2 + \dot \phi_{\vec r}^2  ) \right) \\
&+ \frac{\kappa}{2} \sum_{\vec r} \sum_{l,m=1}^2  s_{lm, \vec r}^2 ,
\end{split}\end{equation}
where $x_{l\vec r}$ denotes the displacement in the $\hat e_{l}$ direction for the mass point in the unit-cell labeled by $\vec r$, and $s_{lm,\vec r}$ denotes the spring extension of the $(l,m)$-th spring in the unit cell, as indicated in Fig.~\ref{fig:aModel}a.
As in the main text, we linearize $s_{lm, \vec r}$ and obtain the rigidity matrix via $\vec S = R \vec X + \mathcal O(\vec X^2)$, where $\vec X$ is the collective vector for the small fluctuations $\{x_{\vec r 1},x_{ \vec r 2},r \delta  \theta_{\vec r}  , r \delta \phi_{\vec r} \}$ with $\delta \theta_{\vec r} = \theta_{\vec r} - \theta_0$ and $\delta \phi_{\vec r} = \phi_{\vec r} - \theta_0$. After Fourier transform, one finds
\begin{widetext}
\begin{equation}\label{eq:}
R_{\vec k} = 
\ell^{-1}
\left(
\begin{array}{cccc}
- a - r \sin \theta_0 & - b - r \cos \theta_0 & a \cos\theta_0 - b \sin \theta_0 & 0\\
( 1 - a - r \sin \theta_0 )e^{-i k_1} & -( b + r \cos \theta_0 )e^{-i k_1}  & - (1-a) \cos \theta_0 - b \sin \theta_0 & 0\\
- b - r \cos \phi_0& -a - r \sin \phi_0 & 0 & a \cos \phi_0 - b \sin \phi_0\\
-(b + r \cos \phi_0) e^{-i k_2} & (1-a - r \sin \phi_0) e^{-i k_2} & 0 & -(1-a) \cos \phi_0 - b\sin  \phi_0
\end{array}
\right),
\end{equation}
\end{widetext}
where $\ell = \text{diag}(\ell_{11},\ell_{12},\ell_{21},\ell_{22})$.

As in the main text, the phase of the system is encoded in the winding numbers of the corresponding fermionic Hamiltonians $\mathcal H_{F}(k_2)$, which we write as
\begin{equation}\begin{split}\label{eq:}
\det\left( i R_{ \vec k}\right) = &
\left(\tilde v_1 + \tilde v_2\,  e^{- i k_2} \right) + \left(\tilde w_1 + \tilde w_2\,  e^{- i k_2} \right)  \, e^{- i k_1}\\
\equiv & \tilde r_1(k_2) + \tilde r_2(k_2) \, e^{- i k_1} .
\end{split}\end{equation}
The winding number $\mathcal W_{k_2}$ is determined by $| \tilde r_1(k_2)| / | \tilde r_2(k_2)|$. After some algebra, one sees that it is determined by the sign of 
\begin{equation}\begin{split}\label{eq:}
\tilde f(k_2) =& (\tilde v_1^2+\tilde v_2^2 - \tilde w_1^2 - \tilde w_2^2) + 2 (\tilde v_1 \tilde v_2 - \tilde w_1 \tilde w_2) \cos(k_2)\\
\equiv & \tilde f_1 + \tilde f_2 \cos(k_2),
\end{split}\end{equation}
where $\tilde f_1$ and $\tilde f_2$ are both real.
More concretely, $\mathcal W_{k_2} =  0 ~\forall k_2$  if $\tilde f_1 > | \tilde f_2|$;
$\mathcal W_{k_2} = -1 ~\forall k_2$  if  $\tilde f_1 < - | \tilde f_2|$; and $\mathcal W_{k_2\rightarrow 0+} \neq \mathcal W_{k_2 \rightarrow \pi^-}$ otherwise, indicating the system is in the TNS phase. For the first two cases, $\mathcal W_{k_2}$ is independent of $k_2$ and it is meaningful to define the integer $n_1 = - \mathcal W_{k_2}$.

The same analysis can be performed with the role of $k_1$ and $k_2$ interchanged, giving the winding number $\mathcal W_{k_1}$. When the phonon spectrum is gapped,  the topological polarization $\vec P_{T} =   n_1 \hat e_1 + n_2 \hat e_2$ is well defined~\cite{Kane2014, Paulose2014}. We plot in Fig.~\ref{fig:aModel}b an example phase diagram of the system, demonstrating all the mentioned phases with different topological polarizations can be accessed in this model.

\end{appendix}

\bibliographystyle{apsrev}   
\bibliography{TP_Refs}

\end{document}